# Boundary-localized many-body bound states in the continuum

Na Sun[*], Weixuan Zhang[*, +], Hao Yuan, and Xiangdong Zhang[$]

*Key Laboratory of advanced optoelectronic quantum architecture and measurements of Ministry of Education, Beijing Key Laboratory of Nanophotonics & Ultrafine Optoelectronic Systems, School of Physics, Beijing Institute of Technology, 100081, Beijing, China*

*These authors contributed equally to this work.

[$,+]Author to whom any correspondence should be addressed: zhangxd@bit.edu.cn; zhangwx@bit.edu.cn

**Bound states in the continuum (BICs), referring to spatially localized bound states with energies falling within the range of extended modes, have been extensively investigated in single-particle systems, leading to diverse applications in photonics, acoustics, and other classical-wave systems. Recently, there has been theoretical interest in exploring many-body BICs in interacting quantum systems, which necessitate the careful design of impurity potentials or spatial profiles of interaction. Here, we propose a type of many-body BICs localized at boundaries, which can be purely induced by the uniform onsite interaction without requiring any specific design of impurity potential or nonlocal interaction. We numerically show that three or more interacting bosons can concentrate on the boundary of a homogeneous one-dimensional lattice, which is absent at single- and two-particle counterparts. Moreover, the eigenenergy of multi-boson bound states can embed within the continuous energy spectra of extended scattering states, thereby giving rise to interaction-induced boundary many-body BICs. Furthermore, by mapping Fock states of three and four bosons to nonlinear circuit networks, we experimentally simulate boundary many-body BICs. Our findings enrich the comprehension of correlated BICs beyond the single-particle level, and have the potential to inspire future investigations on exploring many-body BICs.**

The concept of bound states in the continuum (BICs) was initially proposed by von Neumann and Wigner in 1929 for electrons confined within an inversely designed quantum potential [1]. Subsequently, it was discovered that BICs are a general phenomenon applicable to both quantum and classical wave systems [2, 3]. In the context of single-particle systems and their classical counterparts, various mechanisms can give rise to BICs. For instance, certain resonances may be forbidden from coupling with radiation modes due to symmetry or separability constraints, leading to what is known as symmetry-protected BICs or separable BICs [4-8]. Additionally, it is also possible to suppress radiative emissions by carefully tuning a finite number of systematic parameters [9-18]. Notably, classical-wave systems featuring engineered BICs have been found to exhibit extremely high quality-factors and enhanced near-field concentrations. These exceptional properties can offer numerous important applications based on BICs, such as BIC-lasers [11, 19, 20], sensors [21-26], and others [27-34].

On the other hand, in addition to the extensively discussed single-particle systems, it has been demonstrated that BICs can also exist in two-particle lattices through engineering of onsite interactions and impurity potentials at bulk or boundary sites [35, 36]. For instance, the emergence of two-boson BICs localized on the bulk defect is contingent upon the magnitude of impurity potential exceeding that of Hubbard interaction while sharing the same sign [35]. In addition, the Tamm-Hubbard BIC in two-boson models necessitates a specific boundary impurity potential, which possess the opposite sign with the interaction strength [36]. Furthermore, a recent research proposed the quantum statistics-induced two-particle BICs, where a pair of non-interacting anyons can perfectly concentrate on the boundary of a homogeneous one-dimensional lattice with the statistical angle beyond a threshold [37]. Except for the two-particle systems, theoretical predictions have indicated that three-particle BICs can appear as bound pairs localized by standing waves from other particles within finite lattices with spatially modulated on-site interactions [38]. Moreover, investigations have also been conducted on many-body BICs in a one-dimensional Bose-Hubbard chain with an attractive impurity potential [39], expanding upon previously proposed bulk-impurity BICs involving two bosons. Inspired by these prior findings on interaction-induced BICs, it is intriguing to explore mechanisms for generating many-body BICs and experimentally observing their exotic properties.

In this work, we demonstrate that boundary-localized many-body BICs can be solely induced by uniform onsite interactions without the need for engineering impurity potentials or interaction profiles. Through numerical simulations, we show that three or more interacting bosons can localize at the boundary of a one-dimensional lattice model, which cannot be observed in single- or two-boson systems. Additionally, we show that the eigenenergies of these boundary-localized multi-boson bound states can coexist within the continuous energy spectra of extended scattering states, thus giving rise to interaction-induced boundary many-body BICs. Furthermore, by mapping Fock states of three and four bosons onto nonlinear circuit networks, we experimentally simulate these interaction-induced BICs. Our findings enhance the understanding of correlated BICs and may have significant implications for future investigations into interacting BIC phenomena in many-body systems.

**Results.**

**The theory of many-body interaction-induced bound states in the continuum.** We consider a one-dimensional (1D) finite-size lattice sustaining $N$ interacting bosons. In the framework of the standard Bose-Hubbard model, the system Hamiltonian can be expressed as

$$H = -t \sum_{i=1}^{L} (\hat{b}_{i+1}^{+} \hat{b}_i + \hat{b}_i^{+} \hat{b}_{i+1}) + 0.5U \sum_{i=1}^{L} \hat{n}_i (\hat{n}_i - 1), \quad (1)$$

where $L$ is the length of the 1D lattice, $U$ is the on-site interaction between any two bosons, and $t$ is the single-particle hopping strength. $\hat{b}_i$ ($\hat{b}_i^{+}$) and $\hat{n}_i$ correspond to the annihilation (creation) operator and particle number operator at the $i$th site with $[\hat{b}_i, \hat{b}_j^{+}] = \delta_{ij}$ and $\hat{n}_i = \hat{b}_i^{+} \hat{b}_i$ ($i \in [1, L]$), respectively. In this case, the $N$-boson state can be expanded in Fock space as

$$|N; n_1, n_2 \cdots n_L\rangle = \sum_{n_1, n_2, \ldots, n_L = 0}^{n_1 + n_2 + \cdots + n_L = N} \frac{c_{n_1, n_2, \ldots, n_L}}{\sqrt{n_1! n_2! \ldots n_L!}} (\hat{b}_{x_1}^{+})^{n_1} (\hat{b}_{x_2}^{+})^{n_2} \ldots (\hat{b}_{x_L}^{+})^{n_L} |0\rangle, \quad (2)$$

where $|0\rangle$ is the vacuum state and $c_{n_1, n_2, \ldots, n_L}$ is the probability amplitude of finding $n_i$ particle at the $i$th site. By submitting Eq. (2) into Eq. (1), the eigen-equation of $N$-boson states can be obtained.

At first, we focus on the study of three-boson systems with corresponding Fock-space states being expressed as $|N = 3; n_1, n_2 \cdots n_L\rangle$. Three subplots in Fig. 1(a) display spatial distributions of three bosons with different interaction energies. The first one (the blue block) corresponds to the scenario with three non-interacting bosons locating at different lattice sites. The second one (the yellow block) represents the case where two bosons locate at the same site, while the third one is situated at a different site, resulting in an interaction energy of $U$. The last one (the red block) corresponds to three bosons occupying the same lattice site, where the total interaction energy equals to $3U$. These three different types of three-boson states are called as the single-particle scattering states (single-PSSs), two-Boson aggregation states (two-BASs) and three-Boson aggregation states (three-BASs), respectively. By

solving the steady-state Schrodinger equation $H|N=3;n_1,n_2\cdots n_L\rangle = \varepsilon|N=3;n_1,n_2\cdots n_L\rangle$ of three bosons, we can obtain the eigenequation with respect to a Fock-space basis vector $|\ldots0,1,1,1,0\ldots\rangle$ as:

$$\varepsilon c_{|\ldots0,1,1,1,0\ldots\rangle} = -t(c_{|\ldots1,0,1,1,0\ldots\rangle} + \sqrt{2}c_{|\ldots0,0,2,1,0\ldots\rangle} + \sqrt{2}c_{|\ldots0,2,0,1,0\ldots\rangle} + \sqrt{2}c_{|\ldots0,1,0,2,0\ldots\rangle}$$
$$+\sqrt{2}c_{|\ldots0,1,2,0,0\ldots\rangle} + c_{|\ldots0,1,1,0,1\ldots\rangle}). \quad (3)$$

Eigenequations with respect to the Fock-space basis vectors related to two-PSSs and three-PSSs are presented in Supplementary Note 1.

By solving Eq. (3) with the exact diagonalization method, the eigen-spectrum of the three-boson system is obtained, as shown in Fig. 1(b) with $U$=1.7, $t$=1, and $L$=50. Specifically, we can divide the whole eigen-spectrum into three regions based on the spatial profile of three bosons (see Supplementary Note 2 for details). The first energy region, indicated by the blue arrow from $\varepsilon = -6$ to 6, corresponds to the single-PSSs. The second energy region marked by the yellow arrow from $\varepsilon = -0.7$ to 6.4 corresponds to the two-BASs. And, the third energy region, highlighted by the red arrow from $\varepsilon = 6.2$ to 7.5, corresponds to the three-BASs. Intriguingly, there are two additional degenerate three-BASs at $\varepsilon = 5.87$ (enclosed by the red block), which fall outside of the continuous energy band of three-BASs from $\varepsilon = 6.2$ to 7.5. The inset of Fig. 1(b) presents the magnified view of the eigen-spectrum around these two isolated three-BASs marked by red dots, which are embedded into eigenspectra of single-PSSs and two-BASs.

To gain a more comprehensive insight into the distributions of three-boson states, we calculate spatial profiles of the particle-number density $<n_i>$ for eigenenergies within above three distinct regions. Fig. 1(c1) presents the distribution of $<n_i>$ at $\varepsilon = -2.05$, being embedded within the single-PSSs energy region. It is shown that the low-valued particle-number densities exhibit a standing wave profile along the 1D lattice with the sum of $<n_i>$ equaling to 3, resembling the bulk state in the single-particle lattice. In contrast, for the two-BAS at $\varepsilon = 6.29$ and three-BAS at $\varepsilon = 7.47$, the significantly enhanced values of $<n_i>$ appears, as shown in Figs. 1(c2) and 1(c3), respectively. This enhancement arises from the aggregation of multiple bosons occupying the same site. Fig. 1(c4) displays the spatial profile of $<n_i>$ for the three-BAS at an isolated eigen-energy of $\varepsilon = 5.87$. It is evident that the particle-number density exponentially localizes on both endpoints of 1D lattice, corresponding to the scenario with three bosons being confined to lattice boundaries. In addition, it is notable that the eigenenergy of $\varepsilon = 5.87$ also falls within continuous bands of the single-PSSs and two-BAS, suggesting that such boundary-localized three-BAS can be considered as a three-boson BIC. Furthermore, we also calculate the second-order $g^{(2)}(i,j) = \frac{\langle a_i^+ a_j^+ a_i a_j \rangle}{\langle a_i^+ a_i \rangle \langle a_j^+ a_j \rangle}$ and third-order $g^{(3)}(i,j,k) = \frac{\langle a_i^+ a_j^+ a_k^+ a_i a_j a_k \rangle}{\langle a_i^+ a_i \rangle \langle a_j^+ a_j \rangle \langle a_k^+ a_k \rangle}$ correlation functions for the three-body bound state, as shown in Figs. 1(d1) and 1(d2) with $U$=1.7, $t$=1, and $L$=50. Here, $\hat{a}_i$ ($\hat{a}_i^+$) corresponds to the annihilation (creation) operator at the ith site, and $\langle \cdot \rangle$ denotes the expectation value with respect to the eigenstate of three-body BICs. It is clearly shown that the maximum values of second-order and third-order correlation functions appear at $g^{(2)}(1,1)$, $g^{(2)}(L,L)$ and $g^{(3)}(1,1,1)$, $g^{(3)}(L,L,L)$, indicating that three-boson bound states possess a high probability of detecting all bosons at the same boundary site.

To confirm whether the boundary-localized three-BASs correspond to perfect bound states, we numerically calculate the variation of the inverse participation ratio (*IPR*), which is defined as $IPR = \sum_{n_1,n_2,n_3} |c_{n_1 n_2 n_3}|^4$, of all eigenmodes as a function of the lattice length, as shown in Fig. 1(e) with $U$=1.7. It is noted that the *IPR* of a perfect bound state should converge to a fixed value when the lattice length is large enough, while *IPR*s of other leaky modes are always decreased with respect to the increase of the lattice length. It is clearly shown that *IPR*s of two boundary-localized three-BASs (marked by blue points) approach to a saturated value with the lattice size being increased, indicating that those boundary states possess perfect spatial localizations. The splitting of *IPR*s in the short-length region (with $L$<8) is due to the finite size effect, where the localization length is larger than the lattice length. In this case, two boundary-localized three-BASs can couple with each other, making their eigen-energies become separated. In addition, following the other numerical method on the demonstration of BICs [40, 41], we further calculate the scaling of the imaginary part of three-boson eigen-energies by changing the size of the lattice region with artificial losses. In particular, the artificial losses are introduced by adding the non-Hermitian on-site term (0.05*i*) to all lattice sites except for those within the first to $n_l$th lattice sites surrounding each boundary. Fig. 1(f) presents numerical results on the variation of the imaginary part for the three-boson eigen-energies as a function of $n_l$ with $L$=35. We can see that imaginary component of eigen-energies for two boundary-localized three-BASs approaches to zero in an exponential scaling with increasing the size of the lossless region. It is shown that the divergence tendency appears for the three-boson BICs (the blue dots) with $n_l$ reaching to 11. This phenomenon further confirms the existence of boundary-localized three-boson BICs. Thus, two approaches employed in Figs. 1(e) and 1(f) can consistently manifest the perfectly localized nature of boundary many-body bound states.

In addition, the boundary localization of three-boson bound states depends on the interaction strength. In particular, the value of *IPRs* of three-boson bound states exhibits a notable enhancement by increasing the interaction strength, manifesting an enhanced boundary localization. Moreover, the discrete eigen-energies of boundary-localized three-boson bound states tend to move outside the continuous eigen-spectra as the interaction strength increases, giving a threshold of *U*=1.92 for the existence of three-boson BICs (See details in Supplementary Note 3). Beyond this interaction threshold, the eigen-energies of boundary-localized three-boson bound states can move outside the continuous eigenspectra, transforming the interaction BICs into bound states out of the continuum.

Furthermore, it is important to emphasize that both the onsite interaction and boundary effect play crucial roles in the formation of many-body BICs, while no such BICs can be found in a lattice model with periodic boundary conditions. In the case of a lattice model with periodic boundary conditions, all lattice sites are equivalent, thus lacking an effective interface for the emergence of many-body localized states. Additionally, it is crucial to emphasize that the physical origin for the formation of three-boson BICs is distinct from that of previously revealed two-boson counterparts [35, 36]. In Ref. 35, it is shown that the formation of two-particle BICs localized at the bulk lattice site requires particle interaction and *defect potential* at the lattice center. In addition, Ref. 36 has demonstrated that the existence of two-particle boundary BICs requires the boundary effect, particle interaction, and *defect potential* at lattice boundaries. Without the boundary defect potential, the two-particle boundary BICs cannot appear. Differently, when the number of bosons is equal or larger than three, the many-body boundary BICs can exist without the need of boundary defective potential. Such a distinction stems from the existence of three types of three-boson states, the single-PSSs, two-BASs and three-BASs, with different interaction energies. In this case, the two-BASs with the interaction energy being *U* can be regarded as an effective defect potential of three-BASs, making the formation of three-boson BICs do not require any defect potential (See Supplementary Note 4 for details).

Except for the three-boson BICs, the interaction-induced boundary BICs also exist in the system with more than three bosons. To illustrate this phenomenon, we first consider the system with four interacting bosons. Fig. 2(a) displays the variation of *IPRs* for all four-boson eigenstates as a function of the lattice length with *U*=1.1. We can see that there are a pair of four-boson eigen-states possessing saturated *IPR*s with the increase of the lattice length, indicating that these two four-boson eigenstates are bound states. The spatial distribution of $< n_i >$ for the four-boson bound state is depicted in Fig. 2(b), showing a strong boundary concentration with the sum equaling to 4. In addition, eigen-energies of these two bound states are equal to $\varepsilon = 7.9$, lying within four-boson continuous energy bands. These results show that the four-boson states can perfectly localize around the boundary of the 1D lattice and the corresponding eigen-energies are also embedded within the continuum of four-boson bands, indicating the presence of interacting-induced boundary four-boson BICs. Then, we turn to the five-boson system. The variation of *IPRs* for all five-boson eigenstates as a function of lattice length with *U*=0.8 is shown in Fig. 2(c). It is shown that two five-boson bound states with saturated *IPR*s appear (marked by blue dots). Fig. 2(d) presents the distribution of $<n_i>$ for the five-boson bound state. Similar to three- and four-boson cases, the five-boson bound states also show the strong boundary localization. In addition, the eigenenergy of the boundary-localized five-boson state is equal to $\varepsilon = 9.8$, which also falls within the continuous energy bands. Thus, these two five-boson bound states are boundary BICs. Moreover, following the same methods, we also numerically confirm the existence of interaction-induced BICs in six-boson and seven-boson scenarios. Therefore, we believe that interaction-induced BICs can be generalized to more general multi-boson systems. In addition, it is crucial to emphasize that the interaction threshold, corresponding to the case with boundary-localized bound states moving outside the continuous eigen-spectra, is changed for the system with different numbers of bosons. In Fig. 2(e), we numerically present the variation of the interaction threshold as a function of the number of interacting bosons. It is evident that the interaction threshold of multi-boson BICs decreases with rising the number of bosons. In this case, there also exists a particle number threshold that ensures the existence of boundary BICs at the same interaction strength. This is due to the fact that the eigen-energy of a multi-boson BIC is directly proportional to the particle number *N* and interaction strength *U* as $\varepsilon_{\text{BIC}} \sim N(N-1)U/2$. Thus, as the particle number being increased, the threshold of interaction *U* needs to decrease to ensure that the eigen-energy of many-body BICs can still remain within the continuum spectra, which possess the lower growth rate than that of many-body BICs. In addition, the number of these boundary-localized BICs remains unchanged with the increase in the number of particles, and their eigen-energies are degenerated when the lattice length is large enough.

**The experimental simulation of three-boson and four-boson BICs by electric circuits.** Motivated by previous experimental breakthroughs in simulating few-body quantum systems by electric circuits [42-44], optical waveguides [45] and fiber networks [46]. Since directly mapping a nonlinear circuit to a one-dimensional lattice model is challenging, we first map the one-dimensional lattice model to the Fock space. Then, based on the eigen-equations of the basis vectors in the Fock space, we design the circuit's eigen-equations more directly. in this part, we experimentally simulate the interaction-induced boundary BICs of three- and four-boson systems by nonlinear electric circuits. Fig. 3(a) presents the schematic diagram of our designed circuit simulator for three-boson BICs. Specifically, the voltage at a circuit node labeled by $V_{|n_1,n_2,...,n_L\rangle}$ can be considered as the probability amplitude of $c_{n_1,n_2,...,n_L}$ in the mapped Fock space. Circuit nodes with different colors correspond to three-boson Fock states with different interactions, where blue, orange and red nodes denote single-PSSs, two-PASs, and three-PASs, respectively. Capacitors between different circuit nodes are used to simulate effective couplings between two coupled Fock states, where $C_{t1} = C$ corresponds to the effective coupling between single-PSS and two-PAS and $C_{t2} = \sqrt{2}C$ represents the effective coupling between two-PAS and three-PAS. Moreover, we design a self-feedback module to control the effective interaction strength, where circuit nodes with different colors are connected to self-feedback modules with different interaction strengths. As shown in Fig. 3(b), the self-feedback module is composed of a non-linear multiplier (AD633J) and three operational amplifiers (LM6171), which can be used to continually tune the effective interaction strength of the circuit simulator (See Supplementary Note 5 for details). Specifically, we can control the amplification factor of the multiplier (through $R_1$ and $R_2$ shown in Fig. 3(a)) to adjust the effective interaction strength. This allows us to simulate few-boson BICs with different effective interaction strengths. Furthermore, each node is grounded by an inductor $L_g$.

Through the appropriate setting of grounding and connecting, the eigen-equation of the circuit node mapped to a single-PSS can be derived as:

$$(f_0^2/f^2 - 7.6)V_{|...0,1,1,1,0...\rangle} = -\big(V_{|...1,0,1,1,0...\rangle} + \sqrt{2}V_{|...0,0,2,1,0...\rangle} + \sqrt{2}V_{|...0,2,0,1,0...\rangle} +$$
$$\sqrt{2}V_{|...0,1,0,2,0...\rangle} + \sqrt{2}V_{|...0,1,2,0,0...\rangle} + V_{|...0,1,1,0,1...\rangle}\big) + U_{single}V_{|...0,1,1,1,0...\rangle}, \quad (4)$$

where $f$ is the eigenfrequency $(f_0 = 1/2\pi\sqrt{CL_g})$, $V_{|...0,1,1,1,0...\rangle}$ represents the voltage at the circuit node labeled by $|...0,1,1,1,0...\rangle$, and $U_{single}$ is proportional to the multiplication factor of the self-feedback module used to simulate effective interaction strengths of single-particle scattering states. Details for the derivation of eigenequations for other types of circuit nodes (two-PSS and three-PSS) are provided in Supplementary Note 6. It is shown that the eigen-equations with respect to circuit nodes mapped to all basis vectors in Fock space possess the same form, implying the effectiveness of using nonlinear circuits for simulating many-body BICs in the lattice model. In particular, the probability amplitude for the 1D three-boson model $c_{n_1,n_2,...,n_L}$ is mapped to the voltage $V_{|n_1,n_2,...,n_L\rangle}$. The eigenfrequency ($\varepsilon$) is related to the eigenfrequency ($f$) of the circuit in the form of $\varepsilon = f_0^2/f^2 - 7.6$. To experimentally simulate the interaction-induced three-boson BICs, we fabricate the three-boson circuit simulator with the lattice length being $L$=4, where the circuit network contains totally twenty nodes, corresponding to twenty Fock states of three bosons. Fig. 4(c) displays the photograph image for the enlarged view of the fabricated sample with the values of $C$ and $L_g$ being 1 $nF$ and 200 $mH$, respectively. It is shown that three circuit nodes are connected by capacitors $C_{t1} = C$ (boxed in green) and $C_{t2} = \sqrt{2}C$ (boxed in purple), and voltages at these three nodes are defined by $V_{|1110\rangle}$, $V_{|0210\rangle}$ and $V_{|0300\rangle}$. Simultaneously, each circuit node is connected to a self-feedback module (the red dashed box) with an engineered interaction strength. In addition, the tolerance of circuit element is less than 1% to avoid the detuning of circuit responses.

It is noted that the impedance response of a circuit node is related to the local density of state of the corresponding quantum model [42-44, 47-61]. Thus, to test whether interaction-induced three-boson BICs exist at the expected eigen-energy, we measure the frequency-dependent impedance responses of all circuit nodes with the effective interaction strength being $U$=1.7. According to the impedance responses of all circuit nodes at the frequency $f$, we can draw the distribution of the effective particle-number density for the three-boson eigenstate at $\varepsilon = f_0^2/f^2 - 7.6$ with $<n_i> = \sum_{j=1}^{20} |Z_j|^2 n_i(j)$. Here, the sum from $j$=1 to 20 contains all three-boson states $|n_1, n_2, ..., n_L\rangle$. $Z_{j=|n_1,n_2,...,n_L\rangle}$ represents the normalized impedance value of the circuit node labeled by $V_{j=|n_1,n_2,...,n_L\rangle}$, and $n_i(j)$ corresponds to the particle number of the $i$th site for the three-boson state of $j = |n_1, n_2, ..., n_L\rangle$. In Figs. 3(d1)-3(d3), we plot the measured distributions (marked by red points) of the effective particle-number density at 2.92 $kHz$, 2.80 $kHz$, and 5.08 $kHz$. These three frequencies correspond to eigen-energies of the mapped lattice model at $\varepsilon$ =5.87 (three-boson BICs), $\varepsilon$ =6.65 (a three-PAS in the continuous band), and $\varepsilon = -2.51$ (a single-PSS), respectively. At the frequency of 2.92 $kHz$, we can see that the effective particle-number density is maximally localized at two boundaries of the 1D chain, matching to the theoretical prediction (blue points in Fig. 3(c1)). As for other two frequencies, the measured effective particle-number-densities are extended into the bulk region of the lattice chain, being consistent to spatial profiles of the single-PSS and three-PAS. The lower localization strength of the measured BIC can be attributed to the loss effect, arising from imperfect performance of active circuit elements and disorder effects within the circuit sample, which leads to broadening of the measured impedance peaks. In addition, we perform simulations by taking loss effects into account (in particular, we use the AD633JRZ spice model and the parasitic resistance of inductor is 220 ohms), as shown by orange dots in Figs. 3(d1)- 3(d3). It can be seen that the simulation results considering loss effects are consistent with experimental results. Furthermore, as the lattice length increases, the number of active components such as operational amplifiers and multipliers increases rapidly, leading to enhanced nonlinear instability of the system. To further illustrate the influence of lattice length on three-boson BICs, we conducted simulations considering loss effects for three-boson circuits with varying lattice lengths (See Supplementary Note 7 for details). It is shown that distributions of effective particle density in circuits with $L$=4 can also manifest a comparable spatial profile to those observed in longer lattice lengths. Hence, our experimental findings effectively emulate three-boson BICs.

Then, we fabricate a four-boson circuit simulator with $L$=4 and the effective interaction energy being $U$=1.1 (see Supplementary Note 8 for the photo image). Similar to the above three-boson circuit, we measure impedance responses of all circuit nodes to characterize the four-boson BICs. In Figs. 4(a)-4(c), we present the experimental results of effective particle-density distributions at three different frequencies of 2.60 *kHz*, 2.50 *kHz,* and 4.84 *kHz*, which correspond to eigen-energies at $\varepsilon = 7.94$, $\varepsilon = 8.76$ and $\varepsilon = -4.88$, respectively. It is clearly shown that measured effective particle-number density of four bosons is strongly concentrated around lattice boundaries at 2.60 *kHz*, which is matched to the eigen-energy of four-boson BICs. As for other two frequencies, the measured effective particle-densities are extended into the entire lattice chain, corresponding to four-PSS and single-PSS. In addition, the simulation results of nonlinear circuits with and without loss effects are also presented in Figs. 4(a)-4(c), showing a good agreement between measurements and simulations when considering losses.

**Conclusion.**

In conclusion, we have demonstrated the realization of boundary many-body BICs in a one-dimensional lattice. We numerically show that three or more interacting bosons can concentrate on the boundary of a one-dimensional lattice model with homogeneously onsite potential, which cannot be observed in single- or two-boson systems. Moreover, the eigenenergy of boundary-localized multi-boson bound states can embed within the continuous energy spectra of extended scattering states, thereby giving rise to interaction-induced many-body BICs. Furthermore, by mapping Fock states of three and four bosons to nonlinear circuit networks, we experimentally simulate the interaction-induced BICs. Inspired by potential applications of BICs in other classical-wave systems, we expect that BICs in electric circuits could possess potential applications in the fields of electric sensors with high-sensitivity. Moreover, nonlinear circuits are also able to be used as the classical simulators to investigate other quantum few-body phenomena.

**Methods.**

**Sample fabrications and circuit measurements.** We exploit electric circuits by using LCEDA program software, where the PCB composition, stack-up layout, internal layer and grounding design are suitably engineered. The well-designed PCBs are comprised of six layers. Except for the top and bottom two layers, there are three power layers (±15V and +2V) and one grounding layers. Moreover, Copper is poured on both the top and bottom layers for grounding. For the realization of the self-feedback module, we use the operational amplifier LM6171AIMX, AD711JRZ and the multiplier AD633JRZ. In addition, a pair of 2.2 μF tantalum capacitors and 10 μF ceramic capacitors were used in parallel at both ends of the power supply for all operational amplifiers of all PCBs to realize the function of filtering. The operational amplifier and multiplier are supplied by external voltages of ±15V. The reverse input voltage of the subtractor is set to+2V. All capacitors and resistors are selected with suitable parameters and packaging. Moreover, all PCB traces have a relatively large width (0.75 mm) to reduce the parasitic inductance. To ensure the realization of boson BICs in electric circuits, both the tolerance of circuit elements and series resistance of inductors should be as low as possible. Besides, to ensure the tolerance of circuit elements, we use a WK6500B impedance analyzer to select circuit elements with a high accuracy (the disorder strength is only 1%).

In the measurement of impedance spectra, the impedance analyzer WK6500B interfaces were connected to the GND network of the circuit and a circuit node. To ensure experimental stability, the excitation voltage of the impedance analyzer is set to 1V, with a frequency resolution of 1000 points and the scanning speed set to slow. In addition, it is noted that we can directly excite the three-boson BIC in circuits without needing to excite it from the continuum spectrum. In particular, three-body BICs refer to the states where all particles are localized at the boundary, corresponding to Fock-space basis vector of $|3, 0, 0 ... >$ and $| ...,0, 0, 3 >$. We note that the Kirchhoff equation of our designed circuit is in the same form to the three-boson eigen-equation in Fock space, where eigen-frequencies of the circuit correspond to eigen-energies of lattice model and the voltage distribution is mapped to the three-boson eigen-profile in Fock space. Therefore, to excite the BICs in circuits, we only need to excite the effective

Fock-space basis vector $|3,0,0...>$ or $|...,0,0,3>$ by injecting the voltage signal into the circuit node $V_{|3,0,0...\rangle}$ or $V_{|...,0,0,3\rangle}$. Then, by changing the input frequency of the voltage, it becomes evident that the voltage signal reaches its maximum amplitude when resonating with the eigenenergy of three-boson BICs, thereby affirming direct excitation of BICs.

**Data Availability.** All data are displayed in the main text and Supplementary Information. Data files are available from the corresponding author upon reasonable request.

**References.**

**Acknowledgements**. This work is supported by the National Key R & D Program of China No. 2022YFA1404900, Young Elite Scientists Sponsorship Program by CAST No. 2023QNRC001, and the National Natural Science Foundation of China No.12104041.

**Author contributions.** N. Sun and W. Zhang finished the theoretical investigation. N. Sun performed the experiments with the help from H. Yuan. W. Zhang, N. Sun and X. Zhang wrote the manuscript. X. Zhang initiated and designed this research project.

**Competing interests statement.** The authors declare no competing interests.

# Figure captions

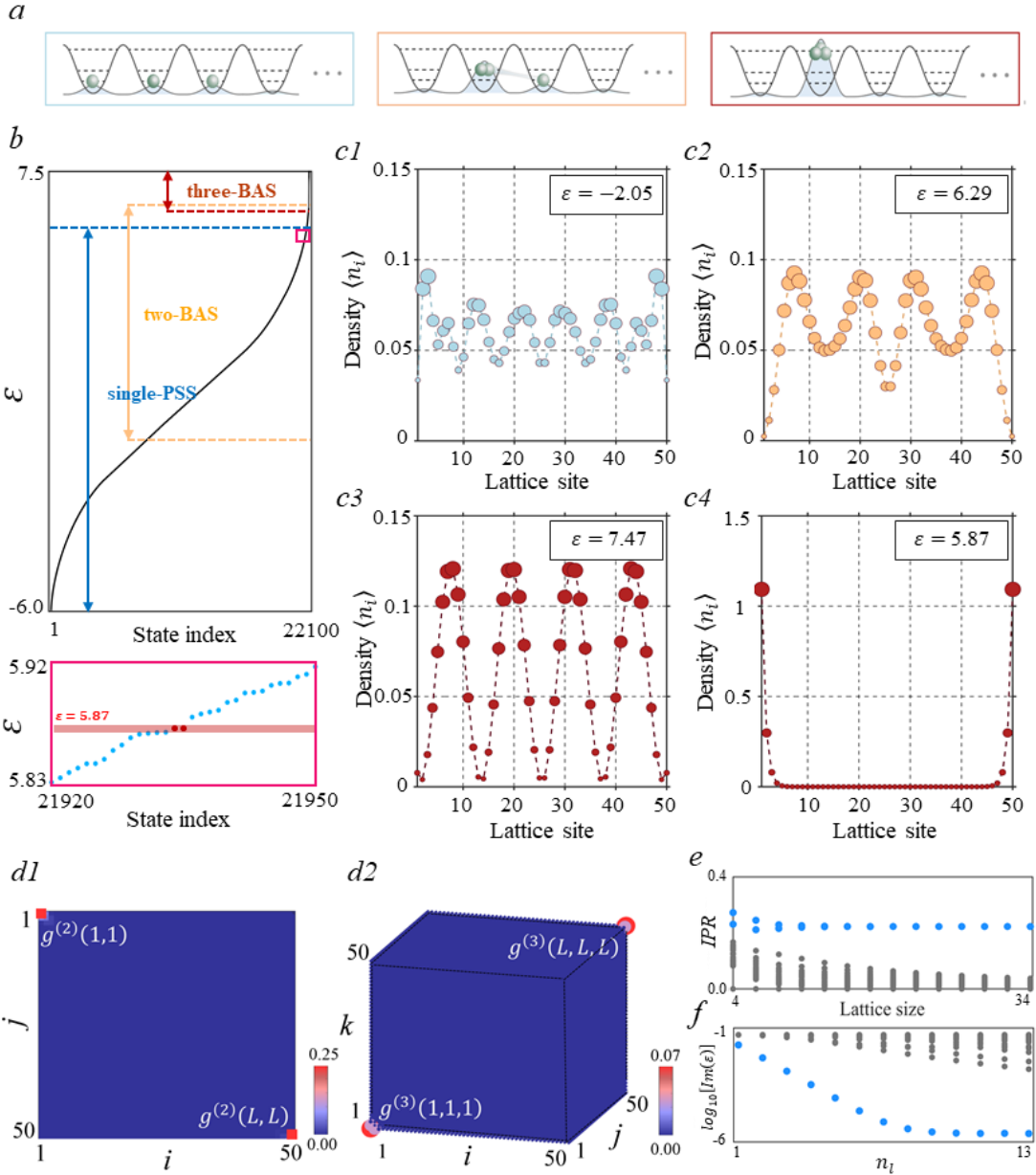

**Figure 1. The theoretical results of interaction-induced boundary BICs of three bosons.** (a). The three cases of spatial distributions of three bosons with different interaction energies. (b). The calculated eigenvalue spectra of the three-boson model with $U$=1.7, $t$=1, and $L$=50. The inset with the magenta frame plots the magnified view of the eigen-spectrum around three-BASs marked by red dots. (c1)- (c4). Numerical results of the particle density distributions for three-boson states at $\varepsilon = -2.05$ (single-PSS), 6.29 (two-BAS), 7.47 (three-BAS), and 5.87 (three-boson BIC), respectively. (d1)- (d2). Numerical results of second-order and third-order correlation functions defined as $g^{(2)}(i,j) = \frac{\langle a_i^+ a_j^+ a_i a_j \rangle}{\langle a_i^+ a_i \rangle \langle a_j^+ a_j \rangle}$ and $g^{(3)}(i,j,k) = \frac{\langle a_i^+ a_j^+ a_k^+ a_i a_j a_k \rangle}{\langle a_i^+ a_i \rangle \langle a_j^+ a_j \rangle \langle a_k^+ a_k \rangle}$ for the three-body BIC. (e). The evolution of IPRs for three-boson eigenmodes as a function of the lattice size with the interaction is 1.7. The blue points indicate the results of IPRs for boundary-localized three-BASs, while the gray points represent the IPRs for other eigenstates. (f). Numerical results on the variation of imaginary parts of eigen-energies as a function of the size for the lossless region with $U$=1.7, $t$=1, and $L$=50. Blue points indicate eigen-energies for two boundary-localized three-BASs, while gray points represent eigen-energies for other states.

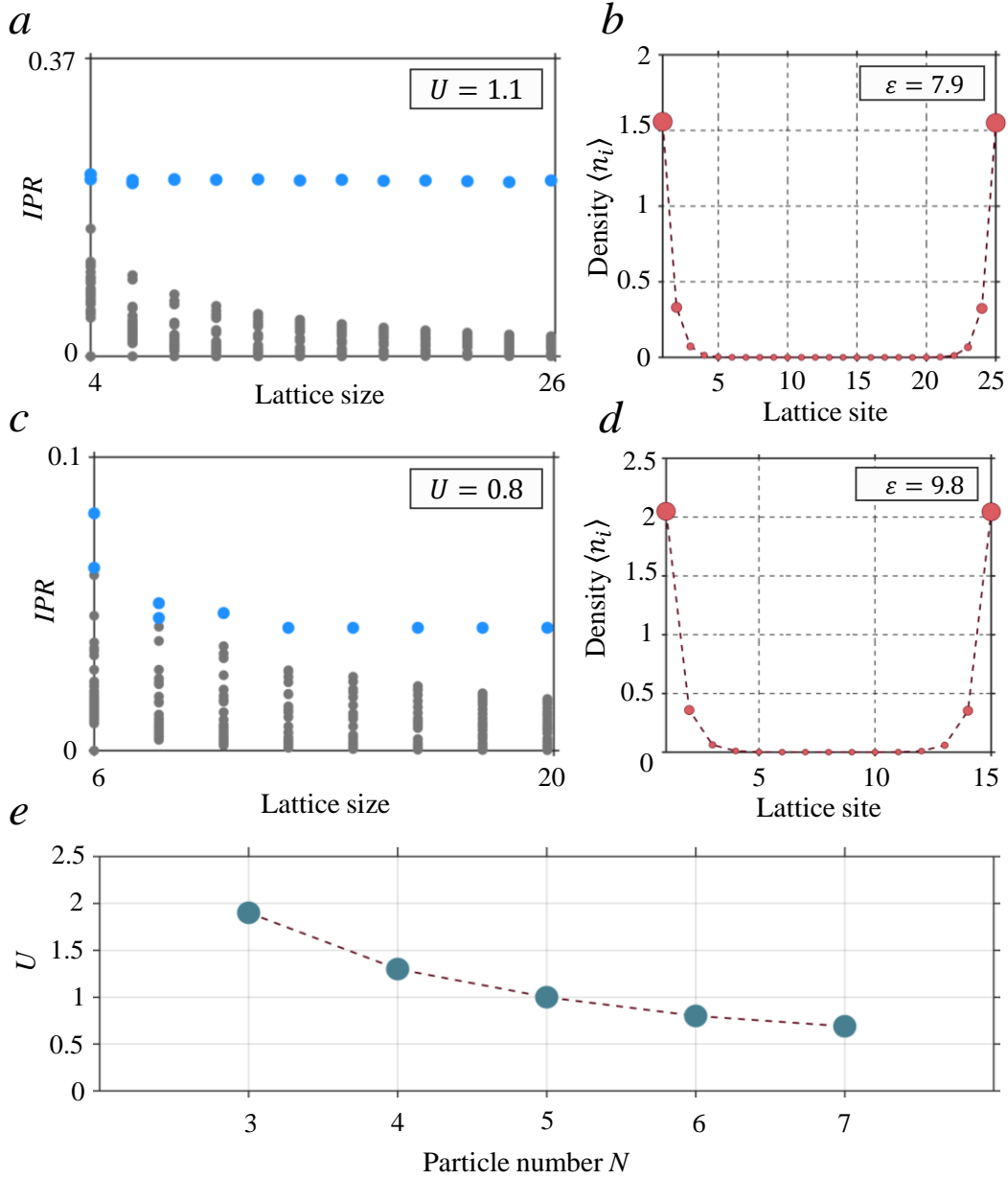

**Figure 2. The theoretical results of boundary many-body BICs.** (a). The evolution of the inverse participation ratio (*IPR*) for four-boson eigenmodes as a function of the lattice size with the interaction is 1.1. Blue points correspond to IPRs for two boundary-localized three-BASs, while gray points represent IPRs for other eigenstates. (b). Calculated the particle density distribution for four bosons state marked by blue. (c). The evolution of *IPR*s for five-boson eigenmodes as a function of the lattice size with the interaction is 0.8. (d). Calculated the particle density distribution for five bosons state marked by blue. (e). Numerical results on the variation of the interaction threshold of BICs as a function of the particle number $N$.

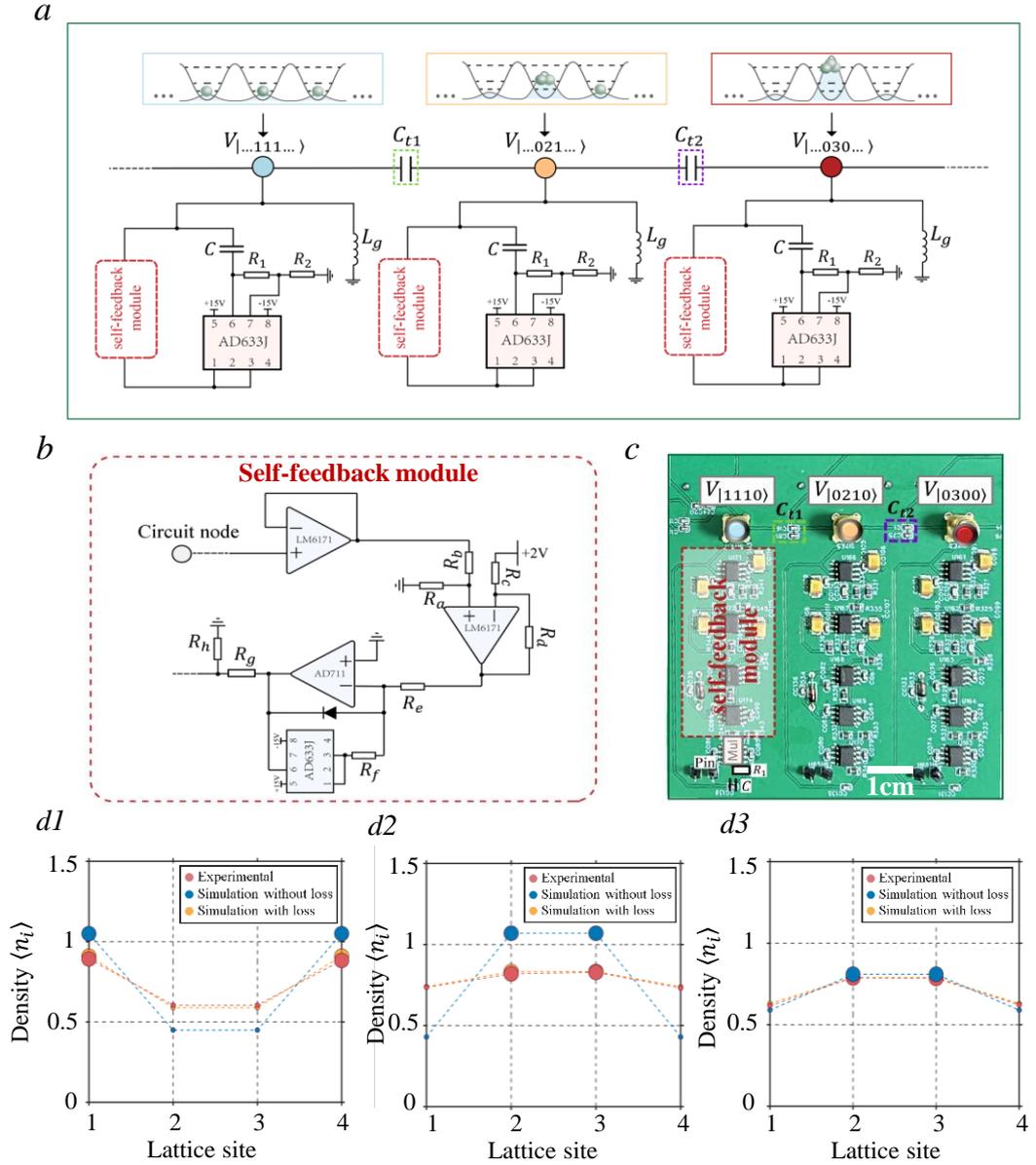

**Figure 3. Experimental results on three-boson boundary BICs by electric circuits.** (a). The schematic diagram for the designed circuit simulator for three bosons. The diagram illustrates the correspondences between three-boson states with voltages. Blue, orange and red nodes represent circuit nodes mapping to three-boson Fock states with different strengths of interaction. Capacitors marked by different types of dotted boxes are used to simulate effective couplings between two Fock states. (b). The scheme of the self-feedback module for controlling the effective interaction strength. The triangle represents the operational amplifier and the square denotes the multiplier. (c). The photograph image for the enlarged view of the fabricated circuit ($L$=4). The voltage signals at three nodes are represented as $V_{|1110\rangle}$ (marked in blue), $V_{|0210\rangle}$ (marked in orange) and $V_{|0300\rangle}$ (marked in orange). (d1)-(d3). Results of effective particle-density distributions at 2.92 *kHz*, 2.80 *kHz*, and 5.08 *kHz*, where red, blue and orange dots present experimental results, simulation results without loss, and simulation results with loss, respectively.

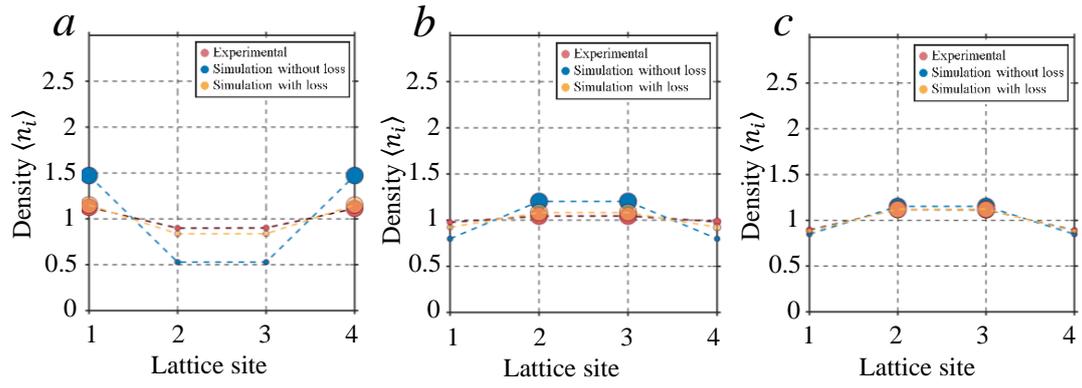

**Figure 4. Measured effective particle-density distributions in the four-boson circuit simulator.** (a)-(c). Red points present the experimental results of the effective particle-density distributions at 2.60 *kHz*, 2.50 *kHz,* and 4.84 *kHz*, respectively. Blue and orange points correspond to simulation results of nonlinear circuits with and without loss effects.